\documentclass[pra,reprint,superscriptaddress]{revtex4-1}

\pdfoutput=1

\usepackage{amsmath,amssymb, graphicx, natbib, mathtools}

\newcommand{\+}{^\dagger}
\newcommand{\nodag}{^{\phantom\dagger}}
\newcommand{\hH}{\hat{H}}
\newcommand{\hy}{\hat{\psi}}
\newcommand{\hty}{\hat{\tilde\psi}}
\newcommand{\kp}{\mathbf{k}_\perp}

\newcommand{\expect}[1]{\langle #1 \rangle}
\newcommand{\ua}{\uparrow}
\newcommand{\da}{\downarrow}
\newcommand{\SF}{SF$_0$}
\newcommand{\NPP}{N$_{\rm PP}$}
\newcommand{\NFP}{N$_{\rm FP}$}

\begin{document}
\author{Bhuvanesh Sundar}
\email{Bhuvanesh.Sundar@uibk.ac.at}
\affiliation{Institute for Quantum Optics and Quantum Information of the Austrian Academy of Sciences, Innsbruck 6020, Austria}
\affiliation{Department of Physics and Astronomy, Rice University, Houston, Texas 77005, USA}
\affiliation{Rice Center for Quantum Materials, Rice University, Houston, Texas 77005, USA}

\author{Jacob A. Fry}
\email{jaf15@rice.edu}
\affiliation{Department of Physics and Astronomy, Rice University, Houston, Texas 77005, USA}
\affiliation{Rice Center for Quantum Materials, Rice University, Houston, Texas 77005, USA}

\author{Melissa C. Revelle}
\email{mrevell@sandia.gov}
\affiliation{Sandia National Laboratories, Albuquerque, New Mexico 85185, USA}

\author{Randall G. Hulet}
\email{randy@rice.edu}
\affiliation{Department of Physics and Astronomy, Rice University, Houston, Texas 77005, USA}
\affiliation{Rice Center for Quantum Materials, Rice University, Houston, Texas 77005, USA}

\author{Kaden R. A. Hazzard}
\email{kaden@rice.edu}
\affiliation{Department of Physics and Astronomy, Rice University, Houston, Texas 77005, USA}
\affiliation{Rice Center for Quantum Materials, Rice University, Houston, Texas 77005, USA}

\title{Spin-imbalanced ultracold Fermi gases in a two-dimensional array of tubes}
\date{\today}

\begin{abstract}
Motivated by a recent experiment Revelle {\em et al}. [Phys. Rev. Lett. {\bf 117}, 235301 (2016)] that characterized the one- to three-dimensional crossover in a spin-imbalanced ultracold gas of $^6$Li atoms trapped in a two-dimensional array of tunnel-coupled tubes, we calculate the phase diagram for this system by using Hartree-Fock Bogoliubov-de Gennes mean-field theory, and compare the results with experimental data. Mean-field theory predicts fully-spin-polarized normal, partially-spin-polarized normal, spin-polarized superfluid, and spin-balanced superfluid phases in a homogeneous system. We use the local density approximation to obtain density profiles of the gas in a harmonic trap. We compare these calculations with experimental measurements in Revelle {\em et al.} as well as previously unpublished data. Our calculations qualitatively agree with experimentally measured densities and coordinates of the phase boundaries in the trap, and quantitatively agree with experimental measurements at moderate-to-large polarizations. Our calculations also reproduce the experimentally observed universal scaling of the phase boundaries for different scattering lengths at a fixed value of scaled intertube tunneling. However, our calculations have quantitative differences with experimental measurements at low polarization and fail to capture important features of the one- to three-dimensional crossover observed in experiments. These suggest the important role of physics beyond-mean-field theory in the experiments. We expect that our numerical results will aid future experiments in narrowing the search for the Fulde-Ferrell-Larkin-Ovchinnikov phase.
\end{abstract}
\maketitle

\section{Introduction}
The Fulde-Ferrell-Larkin-Ovchinnikov (FFLO) phase is a superfluid phase of matter which was originally predicted to occur in superconductors under high magnetic fields~\cite{fulde1964superconductivity, larkin1965nonuniform}. It is unique in that both superconductivity and magnetism coexist in this phase---superconductivity arises from the usual pairing of fermions, while magnetism arises from a net spin induced by the Zeeman effect~\cite{radzihovsky2010imbalanced, kinnunen2018fulde, casalbuoni2004inhomogeneous}. The experimental observation of the FFLO superfluid has been a long-standing challenge.

There are two main difficulties for experimentally realizing the FFLO phase in superconductors under high magnetic fields~\cite{sheehy2006bec, sheehy2007bec, parish2007finite}. First, when a magnetic field is applied to a superconductor, the Meissner effect occurs -- the magnetic field is expelled by induced currents, up to a critical field. Therefore, no net spin is induced. Beyond the critical field, Cooper pairs break due to the large Zeeman energy compared with the superconducting gap. Second, even in the absence of the Meissner effect (such as in either charge-neutral systems or charged two-dimensional systems with an in-plane magnetic field), the parameter space for the FFLO phase is predicted to be small~\cite{sheehy2006bec, sheehy2007bec, parish2007finite, parish2007polarized, radzihovsky2012quantum, casalbuoni2004inhomogeneous}. Despite these difficulties, there is some indirect experimental evidence of FFLO superfluids in two-dimensional organic materials, heavy-fermion materials, and pnictides~\cite{wright2011zeeman, mayaffre2014evidence, koutroulakis2016microscopic, lortz2007calorimetric, beyer2012angle, agosta2017calorimetric, bianchi2003possible, matsuda2007fulde, cho2017thermodynamic, ptok2013fulde, ptok2014influence, ptok2015multiple, zocco2013pauli}.

The FFLO phase can also be potentially realized in other experimental scenarios, such as a two-component fermionic system with a mass imbalance~\cite{sun2013pair, chung2017multiple, parish2007polarized, iskin2006two, mathy2011trimers, lydzba2020unconventional}, atomic Fermi gases at unitarity~\cite{yoshida2007larkin, bulgac2008unitary, bulgac2012unitary, frank2018universal, gubbels2013imbalanced, blume2008trapped}, with spin-orbit coupling~\cite{xia2015three, jiang2014spin, hu2014gapless, dong2013fulde, zheng2014fflo}, in superconducting rings~\cite{yanase2009angular, ptok2012fulde}, and in electron-hole bilayers~\cite{parish2011supersolidity}. A recent theoretical work~\cite{inotani2020radial} argues that the FFLO phase can be realized in vortices in a spin-imbalanced three-dimensional (3D) Fermi gas, a scenario that has previously been realized experimentally~\cite{zwierlein2006fermionic}. A FFLO-like phase is predicted to occur in dense quark matter~\cite{alford2008color} and nuclear matter~\cite{muther2003phases}. But so far, the definitive experimental proof of the FFLO state -- the observation of nonzero pair momentum -- has not yet been attained.

Ultracold atomic gases, which are charge-neutral, are ideally suited to directly probe the presence of the FFLO phase, circumventing some of the limitations of the condensed-matter experiments. Due to the experimental ability to control the initial spin polarization via radiofrequency sweeps, one can potentially realize the FFLO phase in the way it was originally envisioned by Refs.~\cite{fulde1964superconductivity, larkin1965nonuniform}, i.e., in spin-imbalanced fermionic systems, without competing with the Meissner effect that occurs with magnetic fields in charged systems. {\em In situ} imaging in cold atom experiments potentially allows researchers to directly probe the coexistence of magnetism and superfluidity, and the harmonic trapping potential enables measurements of the phase diagram over a wide range of densities. Confining atoms in quasi-one-dimensional (1D) tubes enlarges the parameter space with the FFLO phase as the ground state. Cold atom experiments can also, in principle, implement other experimental scenarios described above to realize the FFLO phase by trapping different atomic species with different masses, by tuning the interaction to unitarity via a Feshbach resonance, by inducing artificial spin-orbit coupling by using Raman lasers, or by trapping them in ring geometries.

One of the most promising steps towards observing the FFLO phase was in a spin-imbalanced $^6$Li gas trapped in a two-dimensional (2D) array of tunnel-coupled quasi 1D tubes~\cite{revelle20161d, revelle2016quasi}. These experiments found that the harmonic trap separates the gas into fully-spin-polarized, partially polarized, and unpolarized phases. Previously, experiments~\cite{liao2010spin} with a 1D gas found density profiles consistent with separation of the trapped gas into FFLO, spin-balanced superfluid, and normal phases, in quantitative agreement with Bethe {\em Ansatz} solutions~\cite{liao2010spin, orso2007attractive}. However, none of these experiments demonstrated superfluidity, provided evidence of domain walls containing the excess $\ua$ atoms, or detected atom pairs with nonzero center-of-mass momentum. Other experiments that have searched for the FFLO phase in spin-imbalanced 2D and 3D atomic gases~\cite{partridge2006pairing, partridge2006deformation, shin2006observation, zwierlein2006fermionic, olsen2015phase, mitra2016phase} have failed to find evidence for it. This is consistent with theoretical predictions that the FFLO phase occupies a very small part of the phase diagram in 2D and 3D gases~\cite{sheehy2006bec, sheehy2007bec, parish2007finite, parish2007polarized, radzihovsky2012quantum, casalbuoni2004inhomogeneous, sheehy2015fulde}.

In this paper, we calculate the phase diagram of a spin-imbalanced Fermi gas trapped in a 2D array of tunnel-coupled 1D tubes by using Hartree-Fock Bogoliubov-de Gennes (BdG) mean-field (MF) theory, over a broad range of experimentally relevant parameters, including those in Ref.~\cite{revelle20161d} and additional measurements presented here. We use the local density approximation (LDA) to calculate the density profiles of both spins in a harmonically trapped gas, as well as the phase boundaries of the gas in the trap, and compare these to experimental measurements. Our calculations qualitatively agree with the measured density profiles, and also reproduce the experimentally observed universal scaling of the measurements when the tunnel coupling is scaled by the pair binding energy.

Although several previous theoretical works~\cite{parish2007quasi, orso2007attractive, liao2010spin, liu2007fulde, hu2007phase, guan2007phase, lee2011asymptotic, yang2005realization, mizushima2005direct, patton2017trapped, patton2020hartree, tezuka2008density, batrouni2008exact, feiguin2007pairing, wei2018fulde, rizzi2008fulde, koponen2008fflo, kim2012fulde, heikkinen2014nonlocal, koponen2007finite, loh2010detecting, pilati2008phase, gubbels2008renormalization, liu2007mean, dutta2016dimensional, mathy2011trimers, parish2013highly, hu2006mean, son2006phase, baksmaty2011concomitant, jensen2007non, yang2013quantum, wolak2012pairing, toniolo2017larkin, rizzi2008fulde, wei2018fulde, zhao2008theory, lin2011u, yang2005realization, mizushima2005direct, patton2017trapped, patton2020hartree, sheehy2015fulde, rosenberg2015fflo, chiesa2013phases, wang2020superfluidity2, pkecak2020signatures, chen2020unusual} have calculated the phase diagram of spin-imbalanced fermions in different scenarios, new calculations are needed to directly compare with the recent measurements~\cite{revelle20161d}. Researchers have calculated the phase diagram in the limit of uncoupled 1D tubes by using exact methods like the Bethe {\em Ansatz} or exact diagonalization for small systems~\cite{orso2007attractive, liao2010spin, liu2007fulde, hu2007phase, guan2007phase, lee2011asymptotic, he2009magnetism, pkecak2020signatures}, DMRG~\cite{feiguin2007pairing, wei2018fulde, rizzi2008fulde, tezuka2008density}, quantum Monte Carlo~\cite{batrouni2008exact}, pairing fluctuation theory~\cite{wang2020superfluidity2, wang2020superfluidity1, chen2020unusual}, as well as approximate methods like MF theory~\cite{mizushima2005direct, patton2017trapped, patton2020hartree}. While exact methods like quantum Monte Carlo are sometimes used for calculating the phase diagram in higher dimensions, too~\cite{wolak2012pairing}, MF theory is the commonly used method, which researchers have used to calculate the phase diagram for a 2D gas~\cite{parish2013highly, toniolo2017larkin, sheehy2015fulde, chiesa2013phases}, a 3D gas with no lattice~\cite{pilati2008phase, gubbels2008renormalization, liu2007mean, dutta2016dimensional, parish2007polarized, sheehy2007bec, sheehy2006bec}, a 3D gas with a 3D lattice~\cite{koponen2008fflo, kim2012fulde, heikkinen2014nonlocal, koponen2007finite, loh2010detecting, rosenberg2015fflo}, and in the polaron limit of large spin imbalance~\cite{mathy2011trimers, parish2013highly}. The phase diagram of a 3D gas with a 2D lattice of tubes, which is the trapping geometry in the experiments we consider~\cite{revelle20161d}, was calculated in Ref.~\cite{parish2007quasi, ptok2017influence} by using MF theory, and in Refs.\cite{zhao2008theory, lin2011u} by using a perturbative treatment away from the exact solution for an uncoupled tube. However, they did not calculate density profiles, a sufficiently broad regime of the phase diagram, or other experimental observables such as spatial coordinates of phase boundaries in a trapped gas, all of which are needed to compare with experiments~\cite{revelle20161d}.

This paper is organized as follows: In Sec.~\ref{sec: setup} we discuss the experimental setup. In Sec.~\ref{sec: MF}, we present the MF theory and the MF phase diagram in a uniform potential. In Sec.~\ref{sec: lda} we use the LDA to calculate the phases and phase boundaries of the gas with harmonic confinement in the axial direction while homogeneous in the transverse directions, and compare these with experimental measurements. We also investigate the universality and 1D-3D crossover observed in experiments. In Sec.~\ref{sec: signature}, we discuss possible experimental signatures of the FFLO phase. We summarize in Sec.~\ref{sec: summary}.

\section{Experimental setup}\label{sec: setup}
\begin{figure}[t]\centering
\includegraphics[width=0.75\columnwidth]{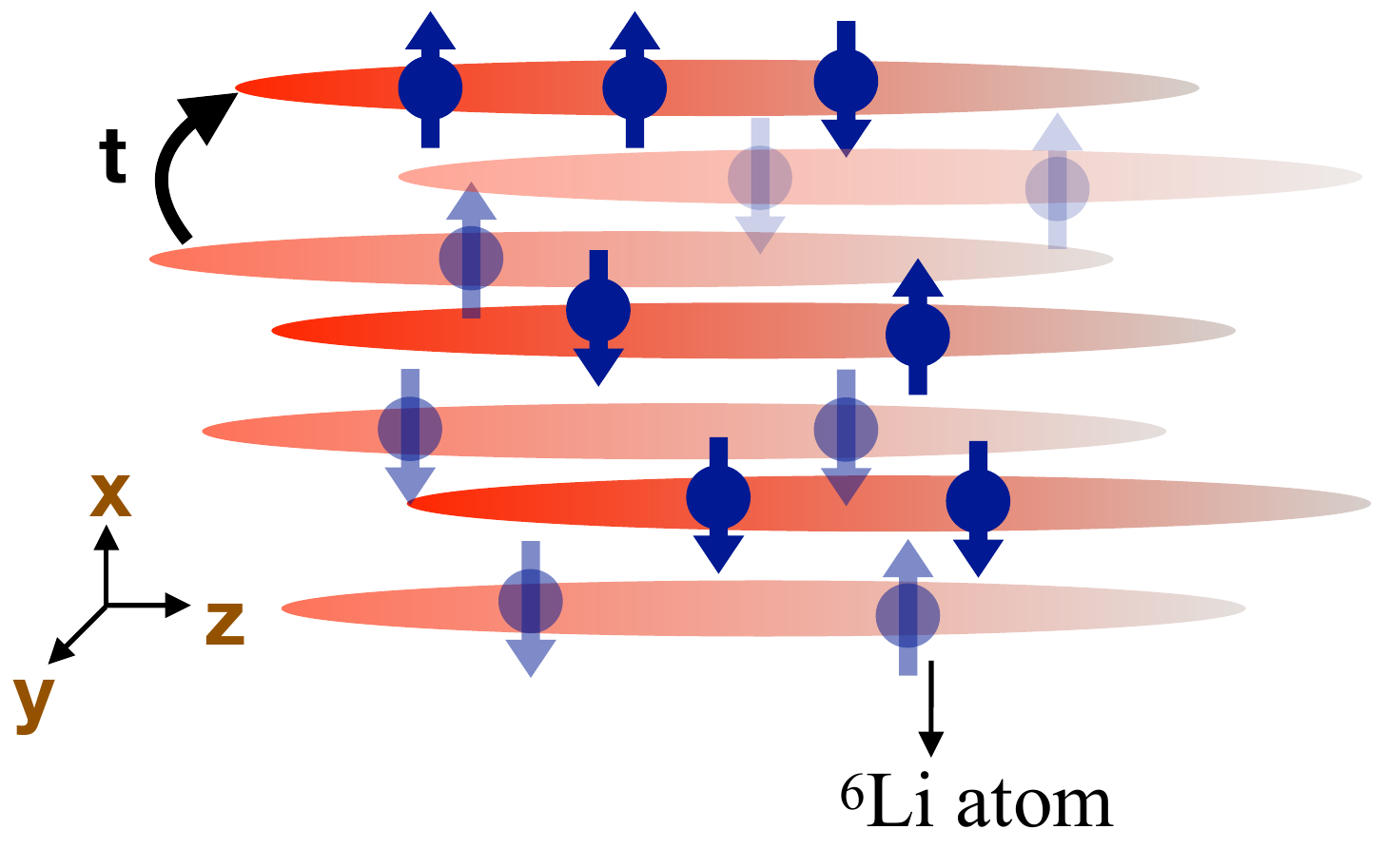}
\caption{(Color online) Schematic of experiment, containing a spin-imbalanced gas of $^6$Li atoms trapped in a 2D array of 1D tubes. The tunneling amplitude between nearest-neighbor tubes is $t$.}
\label{fig: setup}
\end{figure}

\subsection{Setup}

We consider a dilute ultracold gas of $^6$Li atoms trapped in a 2D array of tunnel-coupled 1D tubes along $\mathbf{z}$, as shown in Fig.~\ref{fig: setup}. The tubes are created by a periodic potential, $V(x,y) = V_0\left(\cos^2(\pi x/b)+\cos^2(\pi y/b)\right)$.

The Hamiltonian for the system without harmonic confinement is
\begin{align}
\hH = \int\! d^3\mathbf{r}\ \bigg[& \sum_{\sigma=\ua,\da} \hy\+_\sigma(\mathbf{r}) \left( -\frac{\hbar^2}{2m}\nabla^2 -\mu_\sigma + V(x,y) \right) \hy_\sigma\nodag(\mathbf{r}) \nonumber\\
& + g \hy_\ua\+(\mathbf{r}) \hy_\ua\nodag(\mathbf{r}) \hy_\da\+(\mathbf{r}) \hy_\da\nodag(\mathbf{r}) \bigg].
\end{align}
Here, $\hy_\sigma(\mathbf{r})$ annihilates an atom at position $\mathbf{r}=(x,y,z)$ with spin $\sigma$. The interaction strength $g=4\pi\hbar^2a_s/m$ is parametrized by the 3D scattering length $a_s$, and can be controlled by tuning the magnetic field near a Feshbach resonance. $\mu_\sigma$ is the chemical potential for spin $\sigma$, and can be controlled experimentally via the initial spin populations, which can be set by standard radiofrequency sweep techniques. Spin relaxation is negligible during the experimental duration, and the spin populations remain constant. We assume $\mu_\ua>\mu_\da$, and define
\begin{align}
&\mu = (\mu_\ua+\mu_\da)/2,\nonumber\\
&h=(\mu_\ua-\mu_\da)/2.
\end{align}
We will include the effects of harmonic confinement via the local density approximation in Sec.~\ref{sec: lda}.

\begin{figure}[t]\centering
\includegraphics[width=0.5\columnwidth]{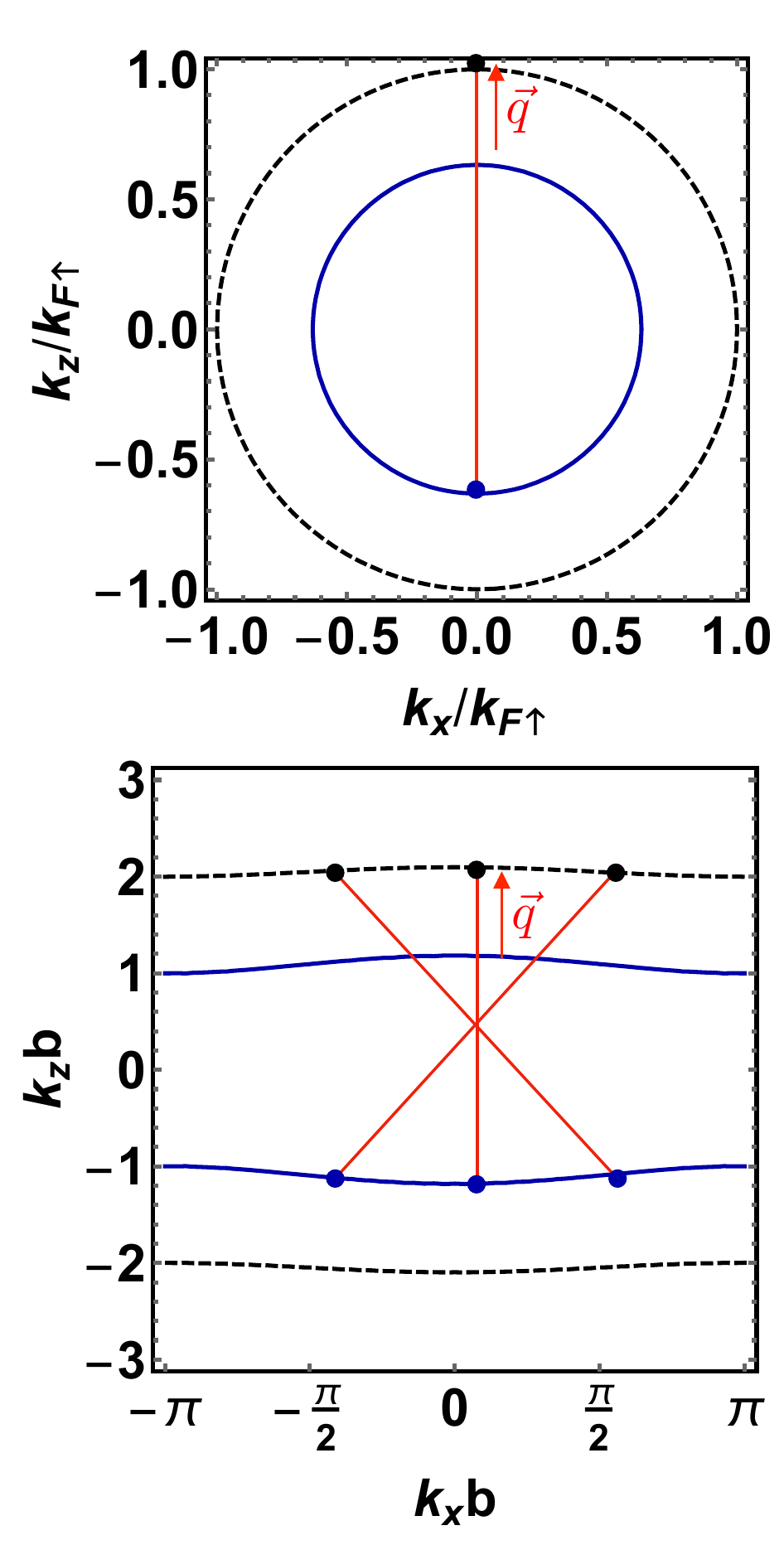}
\caption{(Color online) Illustration of Fermi-surface nesting in an array of tubes, and no nesting in a homogeneous 3D gas. (a) Fermi surfaces for $\ua$ (dashed black) and $\da$ (solid blue) spins in a homogeneous 3D gas, projected onto the $k_y=0$ plane. The momentum difference $\vec{q}$ between the paired $\ua$ and $\da$ spins is unique to each pair. Therefore, there is no nesting, and the FFLO phase is unlikely to be robust. (b) Fermi surfaces for $\ua$ (dashed black) and $\da$ (solid blue) spins in an array of tubes, with $t/E_R = 0.005$, projected onto the $k_y=0$ plane, where $E_R$ is the recoil energy. Multiple pairs of $\ua$ and $\da$ spins on their respective Fermi surfaces have the same momentum difference $\vec{q}$, owing to the nearly flat Fermi surface. Therefore, the Fermi surface is nested, and the FFLO phase is more robust. Paired spins are connected by red lines.}
\label{fig: nesting}
\end{figure}

In the limit where the interaction is weak compared with the lattice band spacing, the system is restricted to the lowest band of the transverse lattice and is well described by a single-band model. In the tight-binding limit where the lattice depth is smaller than the recoil energy, the Hamiltonian becomes
\begin{align}\label{eqn: H}
\hH = \int\! dz & \sum_{\kp} \bigg[ \nonumber\\& \sum_\sigma \hty\+_\sigma(\kp,z) \left( -\frac{\hbar^2}{2m}\partial_z^2 -\mu_\sigma + \epsilon_{\kp} \right) \hty_\sigma\nodag(\kp,z) \bigg] \nonumber\\
& + \frac{g_{1D}}{N_xN_y} \sum_{\mathbf{k_\perp k'_\perp q_\perp}} \hty_\ua\+(\mathbf{k_\perp+q_\perp},z) \hty_\ua\nodag(\mathbf{k_\perp-q_\perp},z) \nonumber\\ & \times \hty_\da\+(\mathbf{k'_\perp-q_\perp},z) \hty_\da\nodag(\mathbf{k'_\perp+q_\perp},z).
\end{align}
Here, $\hty_\sigma(\kp,z)$ annihilates an atom at axial position $z$ and transverse momenta $\kp=(k_x,k_y)$ with spin $\sigma$ in the lowest band of the transverse lattice. $\epsilon_{\kp}=4t - 2t\cos k_xb - 2t\cos k_yb$ is the energy due to tunneling in the $x$ and $y$ directions, where $b$ is the lattice spacing. The sum over $k_x$ and $k_y$ runs over the first Brillouin zone from $-\pi/b$ to $\pi/b$. The effective 1D interaction, $g_{1D}$, is attractive and is related to $a_s$ as~\cite{olshanii1998atomic, bergeman2003atom}
\begin{equation}\label{eqn: EB}
\frac{\sqrt{2} \ell_\perp}{a_s} = -\zeta\left( \frac{1}{2}, \frac{mg_{1D}^2}{8\hbar^3\omega_\perp} \right),
\end{equation}
where $\omega_\perp = \frac{\pi}{b}\sqrt{\frac{2V_0}{m}}$ is the harmonic frequency characterizing the transverse lattice depth, $\ell_\perp = \sqrt{\hbar/m\omega_\perp}$ is the harmonic length in this trap, and $\zeta$ is the Hurwitz zeta function. We denote $\epsilon_B = mg_{1D}^2/4\hbar^2$. This is the 1D pair binding energy. Associated with this energy scale, we define a length scale $\ell_B = \hbar/\sqrt{m\epsilon_B}$.

\subsection{Motivation for this setup}

The motivation for trapping the gas in a 2D array of tunnel-coupled 1D tubes to search for the FFLO phase is illustrated in Fig.~\ref{fig: nesting}. In a 3D gas without the 2D optical lattice, the Fermi surfaces of the $\ua$ and $\da$ spins are spherical, as shown in Fig.~\ref{fig: nesting}(a), and there is no Fermi-surface nesting. Therefore, a 3D gas with no lattice is not favorable for producing the FFLO state. Consistent with this expectation, experiments have so far failed to find any indication of the FFLO phase in a 3D gas with no lattice~\cite{partridge2006pairing, partridge2006deformation, shin2006observation, zwierlein2006fermionic}. In the presence of a 2D lattice in the $x$-$y$ plane, however, the Fermi surfaces are flatter normal to $\mathbf{z}$, as shown in Fig.~\ref{fig: nesting}(b). This leads to large Fermi-surface nesting, and therefore a large parameter space with the FFLO phase as the ground state. Indeed, earlier experiments found experimental signatures in the density profiles that are consistent with the FFLO phase~\cite{liao2010spin}. In the 1D limit, i.e., $t=0$, the Fermi surfaces are two parallel planes for each spin, and are fully nested. But such a 1D gas is not expected to have long-range order. The case of tunnel-coupled 1D tubes, as in Fig.~\ref{fig: nesting}(b), is therefore a promising geometry to search for the FFLO phase, since it potentially combines the large FFLO region of the phase diagram characteristic of 1D with long-range order stabilized by the intertube coupling.

With this motivation, we now move on to calculating the ground state of Eq.~\eqref{eqn: H}, comparing our results with experimental measurements, and provide insight into where the experiments are most likely to find the FFLO phase.

\section{Mean-field theory}\label{sec: MF}

Any eigenstate of Eq.~\eqref{eqn: H} is invariant under the transformations $t\rightarrow \alpha t$, $\mu_\sigma\rightarrow \alpha \mu_\sigma$, $g_{1D}\rightarrow \alpha g_{1D}$ and $z\rightarrow z/\sqrt{\alpha}$, for any constant $\alpha$. Under these transformations, $\epsilon_B\rightarrow \alpha\epsilon_B$ and $\ell_B\rightarrow\ell_B/\sqrt{\alpha}$. Therefore, the phase diagram only depends on the ratios $t/\epsilon_B$ and $\mu_\sigma/\epsilon_B$. We set $\epsilon_B=1$ and $\ell_B=1$, unless otherwise specified.

We make a self-consistent BCS approximation for fermion pairs and a self-consistent Hartree-Fock approximation for the atomic density:
\begin{align}\label{eqn: meanfields}
&\Delta(z) = \frac{g_{1D}}{N_xN_y} \sum_{\kp} \expect{ \hty_\da(-\kp,z) \hty_\ua(\kp,z) } \nonumber\\
& n_\sigma(z) = \frac{1}{N_xN_y}\sum_{\kp} \expect{ \hty_\sigma\+(\kp,z) \hty_\sigma\nodag(\kp,z) },
\end{align}
where $N_xN_y$ is the number of tubes in a finite box with periodic boundary conditions in the $\mathbf{x}$ and $\mathbf{y}$ directions, with each tube having four neighboring tubes. The MF Hamiltonian is
\begin{align}\label{eqn: HMF}
\hH_{\rm MF} =& \sum_{\kp} \int dz \bigg[ \left(\begin{array}{cc} \hy_\ua\+(\kp,z) & \hy_\da\nodag(-\kp,z) \end{array}\right) \nonumber\\
 & \times \left(\begin{array}{cc} H_0 -\mu_\ua+g_{1D}n_\da & \Delta \\ \Delta^* & -H_0 +\mu_\da-g_{1D}n_\ua \end{array}\right)\nonumber\\
 &\times \left(\begin{array}{c} \hy_\ua\nodag(\kp,z)\\ \hy_\da\+(-\kp,z) \end{array}\right)
 - \frac{|\Delta(z)|^2}{g_{1D}} - g_{1D}n_\ua(z)n_\da(z) \nonumber\\ & + \int \frac{dk'}{2\pi} \left(\frac{\hbar^2k'^2}{2m} + \epsilon_{\kp} - \mu_\da + g_{1D}n_\ua(z) \right) \bigg], 
\end{align}
where $H_0 = -\hbar^2\partial_z^2/2m + \epsilon_{\kp}$. To derive Eq.~\eqref{eqn: HMF}, we inserted the mean-field approximations [Eq.~\eqref{eqn: meanfields}] into the Hamiltonian [Eq.~\eqref{eqn: H}], neglected the terms that do not preserve transverse momentum, and used the anticommutation relation $ \hty_\da\+(\kp,z) f \hty_\da\nodag(\kp,z) + \hty_\da\nodag(\kp,z) f \hty_\da\+(\kp,z) = \int dk'/(2\pi) e^{ik'z} f e^{-ik'z}$ for any function $f(z,\partial/\partial z,\cdots)$. The expectation values in Eq.~\eqref{eqn: meanfields} are calculated in the ground state of Eq.~\eqref{eqn: HMF}. We assume a uniform chemical potential and periodic boundary condition along $\mathbf{z}$. In general, $\Delta$ and $n_\sigma$ can depend on $z$, breaking translational symmetry along that direction. The MF approximations made here are expected to be valid as long as $\Delta(z)/\epsilon_B\ll 1$, $g_{1D}n_\sigma/\epsilon_B \ll1$, and for reasonably large $t/\epsilon_B$. When $t/\epsilon_B\ll1$, the system is in the 1D limit, where the quantum fluctuations are large and likely to cause MF theory to fail~\cite{zhao2008theory}.

We numerically find the ground state of Eq.~\eqref{eqn: HMF} that self-consistently satisfies Eq.~\eqref{eqn: meanfields}. To find the ground state, we compare the energy for different self-consistent solutions either obtained analytically or by numerically iterating different initial {\em Ansatz} wave functions as detailed below. We consider a variety of {\em Ansatz} wave functions that are expected to capture all the phases in experiments. The ground state within MF theory is the self-consistent solution with the least energy.

\subsection{Mean-field {\em Ans\"{a}tze}}

We consider the following {\em Ans\"{a}tze}. The {\em Ansatz} for the fully-spin-polarized gas, \NFP, has $n_\da=\Delta=0$ and uniform $n_\ua(z)$. We calculate the solution for this {\em Ansatz} analytically as
\begin{equation}
n_\ua(z) = \frac{1}{N_xN_y}\sum_{\kp} \frac{{\rm Re}[\sqrt{2m(\mu_\ua - \epsilon_{\kp})}]}{\pi\hbar}.
\end{equation}
This solution is self-consistent, i.e., it satisfies Eq.~\eqref{eqn: meanfields}, if $\mu_\da-g_{1D}n_\ua<0$.

The {\em Ansatz} for the partially-spin-polarized normal gas, \NPP, has $\Delta=0$ and uniform $n_\ua(z)>n_\da(z) >0$. We calculate the self-consistent solution for this {\em Ansatz} by solving the implicit equations
\begin{equation}
n_\sigma(z) = \frac{1}{N_xN_y}\sum_{\kp} \frac{{\rm Re}[\sqrt{2m(\mu_\sigma - g_{1D}n_{-\sigma} - \epsilon_{\kp})}]}{\pi\hbar}
\end{equation}
for $n_\ua$ and $n_\da$.

The {\em Ansatz} for the spin-balanced superfluid, \SF, has uniform $n_\ua(z)=n_\da(z)$ and uniform $\Delta(z)\neq0$. We numerically iterate this and all the remaining {\em Ans\"{a}tze}, described below, to self-consistency.

We make two kinds of {\em Ans\"{a}tze} for the FFLO phase -- the FF (Fulde-Ferrell) {\em Ansatz} which has a complex order parameter, and the LO (Larkin-Ovchinnikov) {\em Ansatz} which has a real order parameter.

The FF phase has uniform $n_\ua(z) > n_\da(z) > 0$ and $\Delta(z)=\Delta_0 e^{iqz}$. To obtain the self-consistent solution for the FF phase, we seed the initial {\em Ansatz} with uniform $n_\ua(z)$ and $n_\da(z)$ and $\Delta(z) = \Delta_0 e^{iqz}$, where $n_\ua\ell_B$, $n_\da\ell_B$, and $\Delta_0/\epsilon_B$ for the initial seed are picked randomly from a uniform distribution in the range $[0,1]$. We iterate this {\em Ansatz} to self-consistency by using Eq.~\eqref{eqn: meanfields}. During the iterations after the initial seed, we do not enforce the {\em Ansatz} to be of the form $\Delta(z) = \Delta_0 e^{iqz}$. In an infinite system, the {\em Ansatz} will always retain this form with constant $q$ during the self-consistency iterations, but the form changes in our finite systems, so that the final self-consistent solution is not necessarily the FF phase. We evolve the FF {\em Ansatz} to self-consistency for several values of $q$, and keep the solution with the lowest energy.

The LO phase has $n_\ua(z) > n_\da(z) > 0$ and real $\Delta(z)\neq0$ (except at domain walls), and all three quantities vary with $z$. To obtain the self-consistent solution for this phase, we seed the initial {\em Ansatz} with uniform $n_\ua$, uniform $n_\da$, and $\Delta(z)=\Delta_0(-1)^{\lfloor Mz/L \rfloor}$, where $n_\ua\ell_B$, $n_\da\ell_B$, and $\Delta_0/\epsilon_B$ for the initial seed are chosen randomly from a uniform distribution in the range $[0,1]$, and $M$ is an integer denoting the number of domain walls in the initial seed. We evolve the LO {\em Ansatz} to self-consistency for several values of $M$, and keep the solution with the lowest energy.

When we iterate the LO {\em Ansatz} to self-consistency, two kinds of solutions emerge. In the first kind, the solution has exactly one excess $\ua$ spin per domain wall, i.e., $\int\! dz (n_\ua(z)-n_\da(z))=1$ where the integration region contains one domain wall. This is the commensurate LO phase. In the second kind of self-consistent solution, the solution has a noninteger number of excess $\ua$ spins per domain wall. This is the incommensurate LO phase.

While the FF, commensurate LO, and incommensurate LO are distinct phases, all of them exhibit spin-imbalanced superfluidity, so we group them together as the FFLO phase. We find that the LO phases always have a lower energy than the FF phase. We do not find any other types of solutions besides those described above. We have also checked a large number of the solutions and found that they are robust to different values of the initial seeds.

\begin{figure}[t]\centering
\includegraphics[width=\columnwidth]{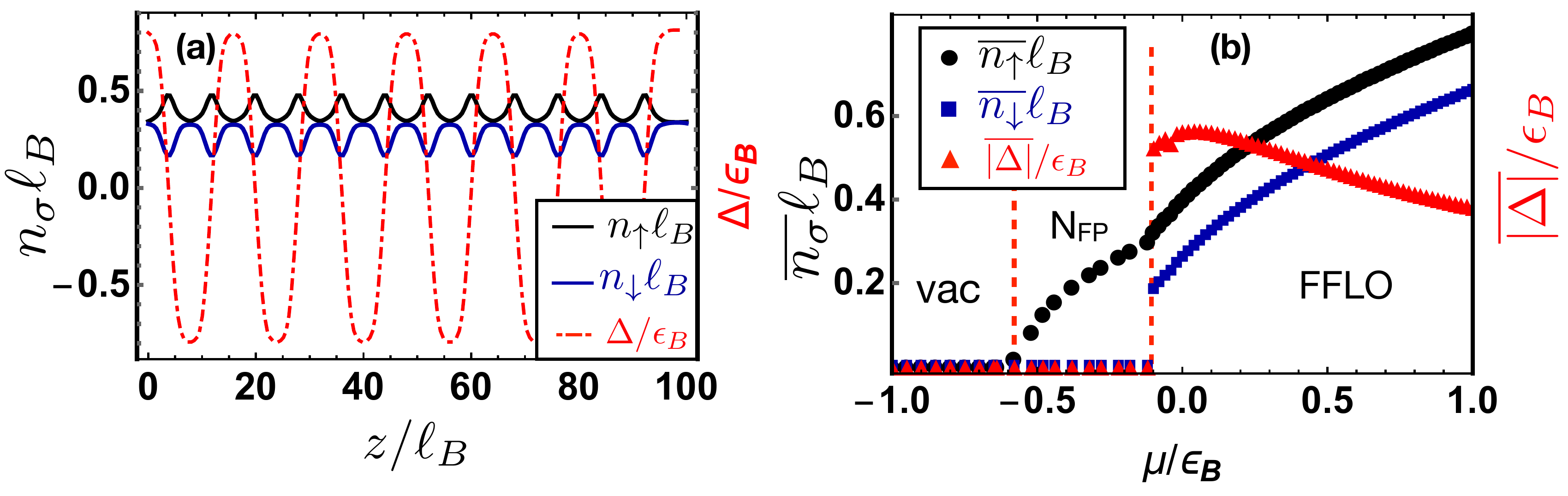}
\caption{(Color online) Dimensionless majority- and minority-spin densities and order parameter in a uniform potential. (a) Densities $n_\ua \ell_B$ (solid black) and $n_\da \ell_B$ (solid blue), and order parameter $\Delta/\epsilon_B$ (dash-dotted red) versus $z/\ell_B$ in a uniform potential with $t=0.02\epsilon_B$, $h=0.63\epsilon_B$, and $\mu=0$. The gas has twelve domain walls in this finite system, and one excess $\ua$ spin per domain wall. This gas is in the commensurate LO phase. (b) Spatially averaged majority- and minority-spin densities $\overline{n_\ua} \ell_B$ (black circles) and $\overline{n_\da} \ell_B$ (blue squares), and spatially averaged order-parameter magnitude $\overline{|\Delta|}/\epsilon_B$ (red triangles) vs $\mu/\epsilon_B$ for $t=0.02\epsilon_B$ and $h=0.63\epsilon_B$. The gas exhibits two phases for these parameters: FFLO at large $\mu$, and \NFP\ at small $\mu$. Red lines separate the different phases.}
\label{fig: density profile 1}
\end{figure}

\begin{figure}[t!]\centering
\includegraphics[width=0.7\columnwidth]{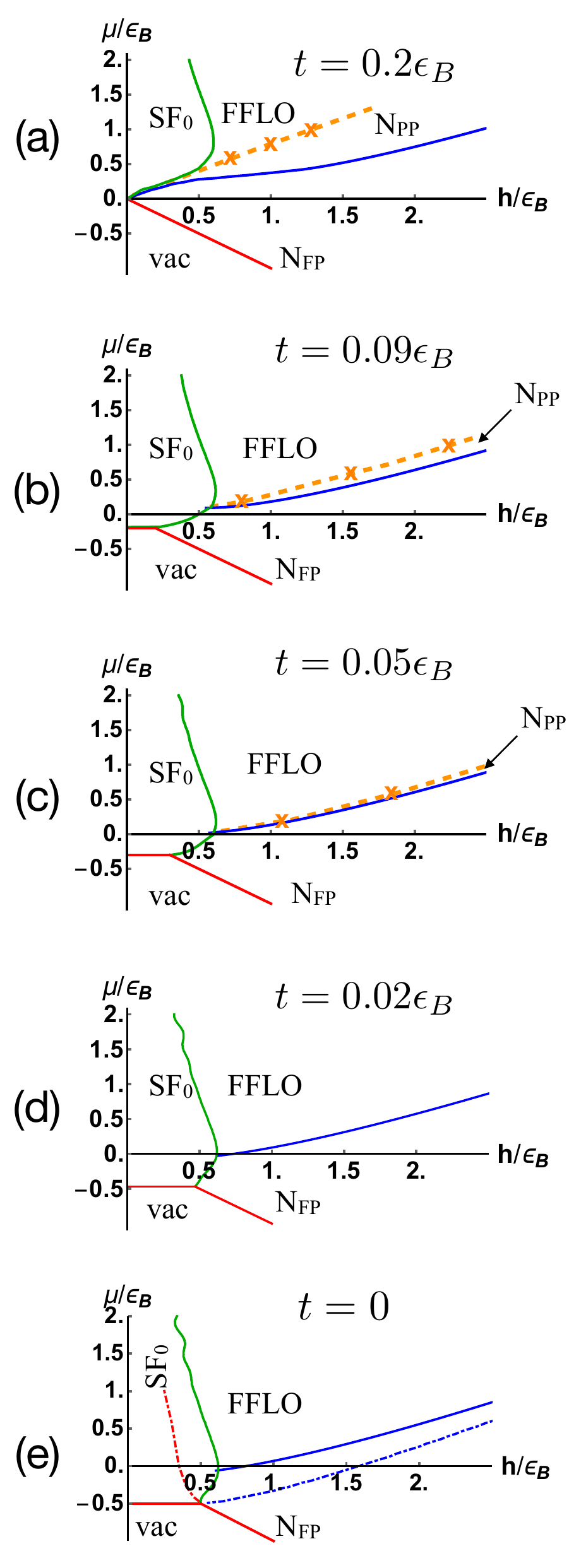}
\caption{(Color online) MF phase diagram of a spin-1/2 Fermi gas in a 2D array of tunnel-coupled 1D tubes, for different tunneling strengths $t$. We observe four different phases: a spin-balanced superfluid (\SF), a spin-polarized superfluid (FFLO), a partially polarized normal phase (\NPP), and fully-spin-polarized normal phase (\NFP). \SF-vacuum and \NFP-vacuum boundaries are in red, \SF-FFLO and \SF-\NFP\ boundaries are in green, and \NPP-\NFP\ and FFLO-\NFP boundaries are in blue. We have only computed a few data points (orange crosses) on the FFLO-\NPP\ boundary, and the dashed orange lines are guides to the eye. Dash-dotted lines in panel (e) show the phase boundaries obtained from the Bethe {\em Ansatz} at $t=0$.
}
\label{fig: phase diags}
\end{figure}

As an example of the self-consistently obtained results, Fig.~\ref{fig: density profile 1}(a) plots $n_\sigma \ell_B$ (black and blue) and $\Delta/\epsilon_B$ (red) versus $z/\ell_B$ in a uniform potential along $\mathbf{z}$, with $t=0.02\epsilon_B, h=0.63\epsilon_B$, and $\mu=0$. To obtain these results, we used a finite system with $L=100\ell_B$ and $N_xN_y=100$ tubes, and discretized space along the axial direction with a grid spacing of $0.5\ell_B$. The order parameter varies with $z$, and has twelve zero crossings, or domain walls, in this finite system of length $100\ell_B$. The spin densities are equal everywhere except near these domain walls, and there is one excess $\ua$ spin at each domain wall. These observations indicate that the gas is in the commensurate LO phase.

From plots like Fig.~\ref{fig: density profile 1}(a), we calculate the spatially averaged value of the spin densities and the order-parameter magnitude $|\Delta|$. The spatially averaged values contain all the information required to determine the phase in a uniform potential. Figure~\ref{fig: density profile 1}(b) plots the spatially averaged spin densities $\overline{n_\sigma} \ell_B$ (black circles and blue squares) and the spatially averaged order-parameter magnitude $\overline{|\Delta|}/\epsilon_B$ (red triangles) in the ground state of a gas in a uniform potential, versus $\mu/\epsilon_B$ at $t=0.02\epsilon_B$ and $h=0.63\epsilon_B$. We find two phases: FFLO for $\mu>-0.1\epsilon_B$ and \NFP\ for $\mu<-0.1\epsilon_B$. There is a discontinuous phase transition from the FFLO to the \NFP\ phase, and the minority-spin density and order parameter changes discontinuously. Repeating this procedure for all $h$ and $t$ gives the full phase diagram.

\subsection{Phase diagram in a uniform potential}\label{sec: bdg}

Figure~\ref{fig: phase diags} shows the system's MF ground-state phase diagram for different tunneling strengths, calculated by using the procedure described above. The parameters for system size and numerical grid spacing are the same as in Fig.~\ref{fig: density profile 1}. There are three or four different phases, depending on the tunneling. The ground state is the \SF\ phase at small $h$ and $\mu>\mu_{\rm vac}$ where $\mu_{\rm vac}$ is a critical value set by $t/\epsilon_B$~\cite{parish2007quasi}. The ground state is the FFLO superfluid at large $h$ and $\mu$. The \NPP\ ground state appears only for $t\gtrsim0.02\epsilon_B$, and occurs at large $h$ and intermediate $\mu$. The ground state is the \NFP\ phase for $h>-\mu_{\rm vac}$, $\mu>-h$ and smaller $\mu$ tha FFLO and \NPP. For $\mu<{\rm min}(\mu_{\rm vac},-h)$, the ground state is the vacuum, which has $n_\ua=n_\da=\Delta=0$.

The phase diagrams in Fig.~\ref{fig: phase diags} are, broadly speaking, qualitatively consistent with experiments~\cite{revelle20161d}, and this will be presented in detail in Sec.~\ref{sec: lda}. Our calculations also distinguish between the FFLO and \NPP\ phases, which have not yet been distinguished from each other by experiments.

The phase diagrams in Fig.~\ref{fig: phase diags} are also consistent with previous MF calculations~\cite{parish2007quasi}, and in rough agreement with the Bethe {\em Ansatz} at $t=0$~\cite{orso2007attractive, liu2007fulde, hu2007phase, guan2007phase}, plotted as dash-dotted lines in Fig.~\ref{fig: phase diags}(e).

Despite the rough agreement, there are two major differences between our results and the Bethe {\em Ansatz}, and one difference between our results and previous MF calculations.

The first difference between MF and the Bethe {\em Ansatz} is the presence of tricritical and multicritical points in the phase diagram. Our phase diagrams have two tricritical points for $t<0.2066\epsilon_B$, consistent with the MF findings in Ref.~\cite{parish2007quasi}. This is in contrast with the Bethe {\em Ansatz} at $t=0$~\cite{orso2007attractive, liu2007fulde, hu2007phase, guan2007phase}, which produces a phase diagram with a multicritical point for four phases instead. Although experiments by using a 2D optical lattice cannot reach $t=0$, they are consistent with having only one multicritical point at $t=0.005\epsilon_B$~\cite{revelle20161d}. This is a failure of MF theory, which is expected since quantum fluctuations become large when the system approaches the 1D limit. As $t$ increases, the tricritical points come closer in MF, and merge at $t=0.2066\epsilon_B$. This is also the tunneling strength where $\mu_{\rm vac}$ reaches $0$~\cite{parish2007quasi}. Current experiments cannot realize such strong tunnelings.

The second difference between MF and the Bethe {\em Ansatz} is the slope of the \SF\ lobe in the $\mu$-$h$ plane. For all $t$, the slope of the lobe is positive from $\mu=\mu_{\rm vac}$, up to a turning point $\mu=\mu_0$ where the slope becomes infinite. As will be discussed in Sec.~\ref{sec: lda}, this implies that a partially-spin-polarized harmonically confined gas can have a \SF\ core, a signature that also occurs in 3D gases with no lattice. Since the positive slope persists up to $t=0$ in MF [see Fig.~\ref{fig: phase diags}(e)], the resulting distribution of phases in the trap is always 3D-like, in the sense of having a \SF\ core in a spin-polarized gas, as long as the central chemical potential is not too large. In contrast, the slope of the \SF\ lobe in the Bethe {\em Ansatz} phase diagram at $t=0$ is negative at all $\mu$, indicating that a harmonically confined gas at any nonzero spin polarization will have a FFLO core. Experiments at $t=0.005\epsilon_B$ are consistent with having a FFLO core at nonzero polarization~\cite{revelle20161d}.

The difference between our results and previous MF calculations~\cite{parish2007quasi} is in the size of the FFLO phase relative to the \NPP\ phase. We find a larger FFLO phase and a smaller \NPP\ phase than Ref.~\cite{parish2007quasi}. This could be because we considered a broader range of {\em Ans\"{a}tze} than Ref.~\cite{parish2007quasi}, which considered only solutions of the FF form to find the boundary between the \NPP\ and FFLO phases. Our results show a shrinking trend for the size of FFLO phase with increasing $t/\epsilon_B$, which is consistent with the expectation that the FFLO ground state is nearly nonexistent in the 3D limit~\cite{radzihovsky2010imbalanced, radzihovsky2012quantum}.

\section{Local density approximation in a harmonic trap}\label{sec: lda}

\begin{figure}[t]\centering
\includegraphics[width=0.7\columnwidth]{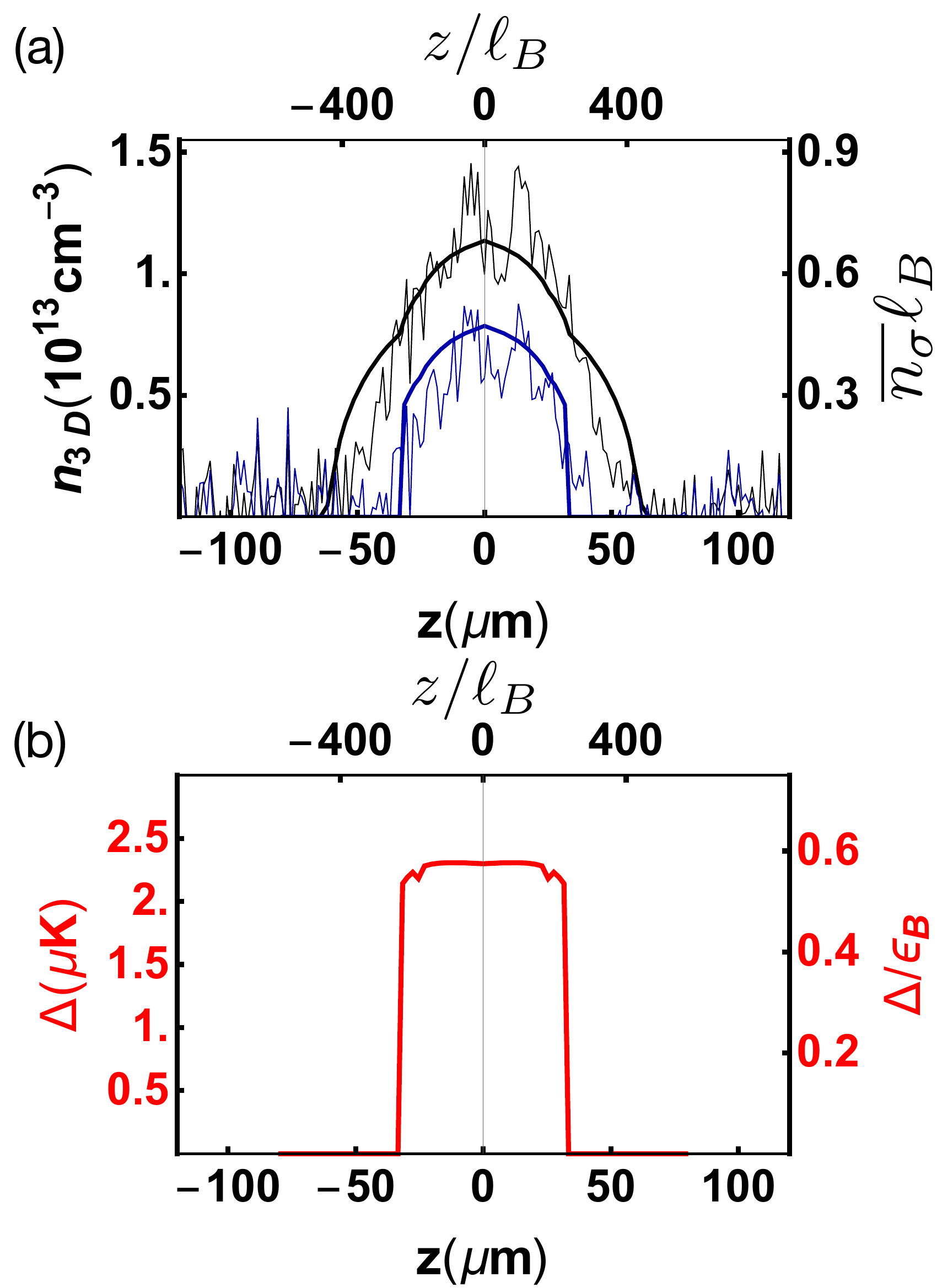}
\caption{(Color online) Majority- and minority-spin densities and order parameter vs axial coordinate in a slowly varying harmonic trap. (a) Spatial profile of the locally averaged 3D spin densities $\overline{n_{\ua,3D}}$ (black) and $\overline{n_{\da,3D}}$ (blue), and (b) locally averaged order-parameter magnitude $\overline{|\Delta|}$ vs $z$, under the local density approximation. The MF curves (thick lines) in panel (a) are overlaid on top of experimental data~\cite{revelle20161d} (thin lines), measured for $t=0.02\epsilon_B$, $N_\ua=268$, $a_s=\infty$, and $P_{\rm tube}=0.4$. These parameters lead to $h=0.63\epsilon_B$ and $\mu_c=0.07\epsilon_B$, which are then used in the LDA to calculate the MF curves. The 3D density is related to the 1D density as $\overline{n_{\sigma3D}}=\overline{n_\sigma}/b^2$. For these parameters, the gas exhibits two phases in the trap: FFLO in the center, and \NFP\ in the wings. The discontinuous phase transition from FFLO to \NFP\ causes a jump in $\overline{n_{\downarrow,3D}}$ and $\overline{|\Delta|}$.}
\label{fig: density profile 2}
\end{figure}

\begin{figure}[t!]\centering
\onecolumngrid
\includegraphics[width=1.0\columnwidth]{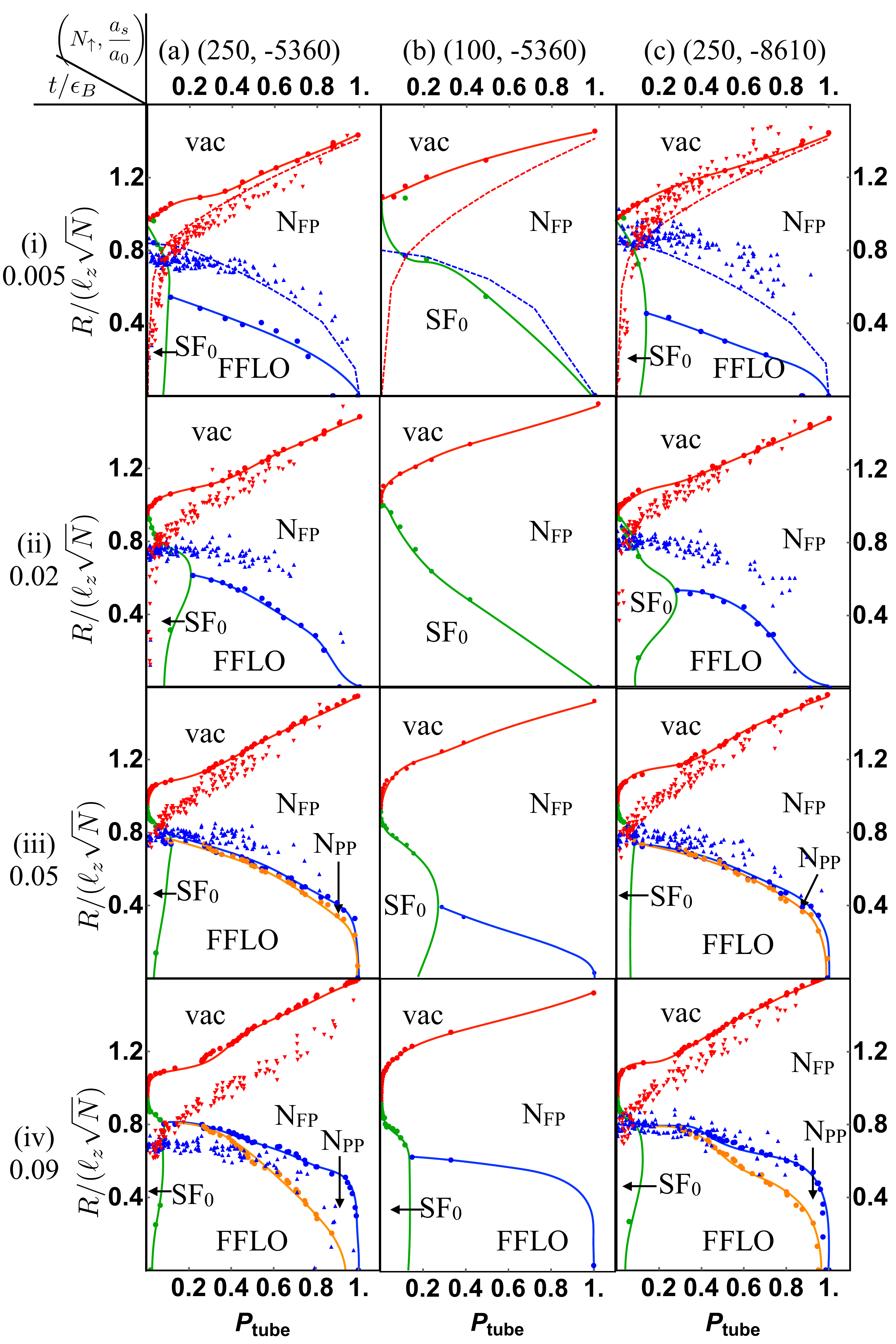}
\caption{(Color online) Scaled radii in the central tube of a spin-1/2 Fermi gas trapped in a 2D array of 1D tubes with axial harmonic confinement, obtained from experiment (triangles) and by using MF theory and the LDA (filled circles) at different tube polarizations $P_{\rm tube}$. The solid lines guide the eye. The dashed lines at $t=0.005\epsilon_B$ are the scaled radii obtained from the Bethe {\em Ansatz}. The parameters for each column are (a) $N_\ua=250, a_s=-5360a_0$, (b) $N_\ua=100, a_s=-5360a_0$, and (c) $N_\ua=250, a_s=-8610a_0$. Parameters for each row are (i) $t=0.005\epsilon_B$, (ii) $t=0.02\epsilon_B$, (iii) $t=0.05\epsilon_B$, (iv) $t=0.09\epsilon_B$. The system exhibits three or four phases: a spin-balanced superfluid (\SF) inside the green lobes, a spin-polarized superfluid (FFLO) for $R<R_\Delta$ (between the orange and green curve), a partially polarized normal phase (\NPP) for $R_\Delta<R<R_\da$ (between the orange and blue curves), and a fully polarized normal gas (\NFP) for $R_\da<R<R_d$ (between the blue and red curves). For some parameters, the \NPP\ phase is not resolvable (if it exists). The blue triangles and red inverted triangles are experimental data points for $R_\da$ and $R_d$.
}
\label{fig: scaled radii}
\twocolumngrid
\end{figure}

In the experiments in Ref.~\cite{revelle20161d}, the 2D optical lattice which creates the array of tubes also results in a slowly varying potential envelope that is approximately harmonic along three axes. As a result of the spatially varying potential, the gas exhibits several phases which appear at different distances from the center. The spin-sensitive {\em in situ} density images partially reveal the phases present in the experiment -- they can distinguish all of the predicted phases except \NPP\ vs FFLO.

In the LDA, the properties of the system at a position $\mathbf{r}$ are approximated to be those of a homogeneous system with chemical potential $\mu(\mathbf{r})=\mu_{\rm vac}-V_{\rm trap}(\mathbf{r})$, where in the present experiments $V_{\rm trap}(\mathbf{r}) = m(\omega_x^2 x^2+\omega_y^2 y^2 + \omega_z^2 z^2)/2$ is a good approximation for the potential. The LDA is accurate in large enough systems. In our experiments, $\omega_z$ varies linearly with $V_0$, from $2\pi\times197$ Hz to $2\pi\times256$ Hz when $V_0$ is varied from $2.5E_R$ to $12E_R$. For these parameters, we expect the LDA to be fairly accurate.

We use the LDA to calculate the variation of densities of both spins and the order parameter along the axial direction in one tube, assuming an axially varying harmonic trap and a uniform potential in the transverse directions. We compare the calculated density profiles with experimentally observed density profiles in one tube. We focus on the density profiles of the central tube in the experiments, which are obtained by doing an inverse Abel transform on the column-integrated densities extracted from images. From the calculated density profiles and the order parameter in the LDA, we extract the regions in the trap where the various phases occur. We can similarly extract the phases from the experimental observations, but cannot distinguish between FFLO and \NPP, because the experiments did not measure the order parameter.

We extract the locally averaged real-space density $\overline{n_\sigma}(z)$ in one tube in MF theory by using the variation of $\overline{n_\sigma}$ with $\mu$ in Fig.~\ref{fig: density profile 1}(a), and the fact that the local potential in the central tube varies in the experiment as $\mu(z) = \mu_c - m\omega^2z^2/2$. We set $\mu_c$ and $h$ as the appropriate chemical potentials which give the right value for the experimentally measured total particle numbers in the central tube for each spin, $N_\sigma = \int_{-\infty}^\infty n_\sigma(0,0,z)\ dz$, which in the LDA are
\begin{equation}\label{eqn: LDA}
N_\sigma = \int_{-\infty}^{\mu_c} \frac{ \sqrt{2}n_\sigma(\mu,h) }{ \sqrt{m\omega^2(\mu_c-\mu)} } d\mu.
\end{equation}
We define the polarization in the tube as
\begin{equation}
P_{\rm tube} = \frac{N_\ua-N_\da}{N_\ua+N_\da}.
\end{equation}
The 3D densities $\overline{n_{\sigma,3D}}$ are related to the 1D densities as $\overline{n_{\sigma,3D}} = \overline{n_\sigma}/b^2$.

Figure~\ref{fig: density profile 2}(a) plots $\overline{n_{\sigma,3D}}$ vs $z$ extracted by doing an inverse Abel transform on the experimental data (thin lines), and the locally averaged densities $\overline{n_{\sigma,3D}}$ obtained from MF (thick lines), for the parameters $t=0.02\epsilon_B, a_s=\infty, N_\ua=268$, and $P_{\rm tube}=0.4$, where the values of $N_\ua$ and $P_{\rm tube}$ are obtained by analyzing the experimental data. For these parameters, we find that $\mu_c=0.07\epsilon_B$ and $h=0.63\epsilon_B$. There is good overall agreement between the experimental and MF density profiles. Similar agreement is observed qualitatively in all the data, although for some parameters, especially those at small polarization, there is up to a $50\%$ difference in the boundaries of the phases in the MF curves and experimental data.

Figure~\ref{fig: density profile 2}(b) plots the locally averaged order-parameter magnitude $\overline{|\Delta|}$ vs $z$. The order parameter is nonzero in the central region of the trap and the gas is spin polarized there, indicating that the phase in the center is FFLO. There is a discontinuous transition to the \NFP\ phase at $z\approx30\mu$m. Experiments have not measured the order-parameter magnitude yet. Figure~\ref{fig: density profile 2}(b) shows that the FFLO phase should be present in a significant region of the trap in experiments. The order-parameter magnitude is $2.3\mu$K in the center of the trap, suggesting that, at least under some conditions, the FFLO state remains robust up to a temperature on this order. The Fermi temperature corresponding to the peak density in Fig.~\ref{fig: density profile 2}(a) is $3\mu$K. The temperature in the experiments is typically well below this.

\subsection{Scaled radii of phase boundaries}

From plots like Fig.~\ref{fig: density profile 2} showing $\overline{n_{\sigma,3D}}$ and $\overline{|\Delta|}$ vs $z$, we extract the axial coordinate of the various phase boundaries. We define $R_d$ as the maximum axial coordinate where $\overline{n_{\ua,3D}}\neq \overline{n_{\da,3D}}$, $R_\da$ as the maximum coordinate where $\overline{n_{\da,3D}}>0$ (which is the inner edge of the \NFP\ phase), $R_{\rm SF,1}$ and $R_{\rm SF,2}$ as the inner and outer edges of the \SF\ phase (where $n_\ua=n_\da>0$), and $R_\Delta$ as the maximum coordinate where $|\Delta|=0$ and $\overline{n_{\ua,3D}}>0$ (which is the outer edge of the \SF\ and FFLO phases combined). Of these, $R_\Delta$ is not measurable experimentally. Some of these coordinates are ill-defined in the limits $P_{\rm tube}=0$ and $P_{\rm tube}=1$. In these cases, the radii are computed or measured in the limit $P_{\rm tube}\rightarrow0^+$ or $P_{\rm tube}\rightarrow1^-$.

We scale the coordinates of the phase boundaries by $\sqrt{N}\ell_z$. This choice of the scaling factor is natural, since the scaled coordinates $R/(\sqrt{N}\ell_z)$ are less dependent on fluctuations in $\sqrt{N}$ and $\omega_z$ in the experiments. This can for example be seen by noting that
\begin{equation}\label{eqn: Rd}
\frac{R_d}{\sqrt{N}\ell_z} = \left( \frac{\hbar}{2m(\mu_c-{\rm min}(\mu_{\rm vac},-h))} \int_{-\infty}^{\mu_c} d\mu \frac{n(\mu,h)}{\sqrt{\mu_c-\mu}} \right)^{-1/2}.
\end{equation}
The right-hand side of this equation does not explicitly depend on $N$ or $\omega_z$. In the special limit $t=0$ and $P_{\rm tube}=1$, all the scaled radii are analytically known; $R_d/(\sqrt{N}\ell_z)=\sqrt{2}$, $R_\da=R_\Delta=0$. In this limit, $\mu_c = N\hbar\omega_z-h$. At small tunnelings at $P_{\rm tube}=1$, the chemical potential can be obtained by Taylor expanding the integral in Eq.~\eqref{eqn: LDA} as $\mu_c = N\hbar\omega_z-h-2t$, leading to
\begin{equation}\label{eqn: perturbative Rd}
R_d/(\sqrt{N}\ell_z)=\sqrt{2(1+4t/N\hbar\omega_z)}.
\end{equation}

\subsection{Scaled radii: Experiment vs Theory}

All the scaled radii described above can be determined by specifying only four parameters: $N_\ua$, $a_s$, $t/\epsilon_B$, and $P_{\rm tube}$. In principle, $\hbar\omega_z/\epsilon_B$ is also a free parameter, but in our calculations as in the experiments, $\omega_z$ is determined from the optical lattice depth which provides the harmonic confinement, and so is not independent of $t$.

The filled circles in Figs.~\ref{fig: scaled radii}(a.i)-\ref{fig: scaled radii}(a.iv) show the scaled radii vs tube polarizations $P_{\rm tube}$, obtained from MF theory for a harmonically trapped gas with various $t/\epsilon_B$ and fixed scattering length $a_s=-5360a_0$ and $N_\ua=250$. The boundaries of the different phases are extracted by using the procedure described earlier in this section. The blue triangles and red inverted triangles show the experimental measurements for the scaled $R_\da$ and $R_d$ in the central tube. The blue, red, green, and orange filled circles are the scaled $R_\da, R_d, R_{\rm SF,i}$, and $R_\Delta$ in MF theory, and the solid lines are guides to the eye. The dashed red and blue lines in Fig.~\ref{fig: scaled radii}(a.i) are the scaled $R_d$ and $R_\da$ obtained from the Bethe {\em Ansatz} at $t=0$. In the Bethe {\em Ansatz}, $R_{\rm SF 1}=R_d$ and $R_{\rm SF 2}=R_\da$ up to the multicritical point at $h=0.5\epsilon_B$, and $R_\Delta=R_\da$ always. Experiments have not yet measured $R_\Delta$ and $R_{\rm SF,i}$. We set the horizontal axis as $P_{\rm tube}$ instead of $h$, since $P_{\rm tube}$ is experimentally observable. Since calculating the scaled radii versus $P_{\rm tube}$ requires us to calculate the self-consistent solution at a large number of points in the $\mu$-$h$ plane for each tunneling, we used a smaller system size of $L=25\ell_B$ with a grid spacing of $0.25\ell_B$. We find the finite-size errors due to the reduced system size to be negligible --- the majority-spin density changed by $O(10^{-3})$ when we reduced our system size from $L=100\ell_B$ to $L=25\ell_B$.

The scaled radii plotted in Figs.~\ref{fig: scaled radii}(a.i)-\ref{fig: scaled radii}(a.iv) are, broadly speaking, qualitatively consistent with the scaled radii derived from experimental data, but there are quantitative differences. At all tunnelings at $P_{\rm tube}=0$, the gas is in the \SF\ phase everywhere in the trap, both in MF theory and the experiment. Here, $R_{\rm SF,1}=0$ and $R_\da=R_\Delta = R_{\rm SF,2}$. However, there is significant difference in the value of the latter scaled radii measured in experiments and obtained from MF theory at $P_{\rm tube}=0$. At all tunnelings at $P_{\rm tube}=1$, the gas is in the \NFP\ phase everywhere in the trap, $R_\da=R_\Delta=0$ and $R_d/(\sqrt{N}\ell_z) \approx \sqrt{2(1+4t/N\hbar\omega)}$ as predicted in Eq.~\eqref{eqn: perturbative Rd}. For large $P_{\rm tube}$, our MF calculations produce a spin-polarized phase in the trap center, surrounded by the \NFP\ phase, agreeing with experimental observations.

Our calculations also reveal phase boundaries that have not yet been measured by experiments, but should be present. For example, the phase boundaries in the MF calculations easily distinguish between FFLO and \NPP; $\Delta(z)\neq0$ in the FFLO phase, and $\Delta(z)$ and $n_\sigma(z)$ vary with $z$ on a length scale given by the difference in Fermi momenta of $\ua$ and $\da$. Our calculations predict that the \NPP\ phase appears for $t>0.02\epsilon_B$ and is either absent or occupies a very small space for $t\leq0.02\epsilon_B$. This is consistent with the Bethe {\em Ansatz} at $t=0$. Previous experiments~\cite{revelle20161d} did not measure $\Delta$, and did not have the spatial resolution to image rapid density oscillations. Therefore, they could not distinguish between FFLO and \NPP. Future experiments which probe the gas after a time-of-flight expansion, or with a high-resolution microscope which can resolve density oscillations {\em in situ}, may be able to distinguish between these two phases~\cite{lu2012expansion, kajala2011expansion}.

The largest disagreement between experiments and our calculations is at small $P_{\rm tube}$ and $t\lesssim 0.02\epsilon_B$, i.e., close to the 1D limit. In the limit $P_{\rm tube}\rightarrow0$, MF theory predicts that $R_d \rightarrow R_\da$, while the Bethe {\em Ansatz} predicts and experiments measure that $R_d\rightarrow0$. Physically, this means that experiments observe a spin-polarized (i.e FFLO or \NPP) phase in the center surrounded by the \SF\ phase, while MF predicts a \SF\ phase in the center surrounded by the FFLO phase and the \NFP\ phase. A distribution of the phases as in experiments for $t\lesssim0.02\epsilon_B$ is often referred to as 1D-like, while the distribution of phases in MF is 3D-like and inverted relative to the 1D-like phase distribution. As the tunneling increases from $t=0$ to $t=0.02\epsilon_B$, experiments observe that the $R_d$ vs $P_{\rm tube}$ curve gets steeper so that the polarization where $R_\da=R_d$ shifts to smaller $P_{\rm tube}$. Beyond $t\approx0.02\epsilon_B$, $R_d=R_\da$ at $P_{\rm tube}=0$, i.e., the experimental measurements are consistent with a 3D-like phase distribution with a \SF\ core in the trap center, agreeing with our MF theory.

In the regime described above where $t\lesssim0.02\epsilon_B$, the experimental measurements agree better with the Bethe {\em Ansatz}. At $t=0$, the Bethe {\em Ansatz}, which is exact, nearly perfectly matches the experimental measurements, with small deviations occurring possibly due to finite-temperature corrections unaccounted for here. At $t=0.02\epsilon_B$, the experiments still observe a 1D-like phase distribution like that at $t=0$.

Some of the differences between experimental measurements and MF calculations could be due to the invalidity of MF theory in some regimes, or systematic inaccuracies in our calculations due to finite system size and finite discretization of our system in space. MF theory is not expected to be valid for $t/\epsilon_B\ll 1$ due to large quantum fluctuations. Indeed, our calculations differ the most from experimental measurements when $t\lesssim0.02\epsilon_B$. In lieu of MF theory, Ref.~\cite{zhao2008theory} proposes a perturbative treatment of the tunneling from an exact solution at $t=0$. Strong interactions could also produce beyond-mean-field effects, or cause a failure of the tight-binding model due to excitation to higher bands in the lattice. We used a finite system size $L=25\ell_B$ with a grid spacing of $0.25\ell_B$ in our calculations in Sec.~\ref{sec: lda}, both of which can contribute to systematic errors. In the FFLO phase at large polarization, the nonzero grid spacing is a source of error because the domain-wall spacing can become comparable to the grid spacing. In the FFLO phase at small polarization, the finite system size is a source of error because the polarization cannot smoothly go to zero in our system. Another source of error could be that $N_\ua$ in the experiments sometimes deviates considerably from $250$, even though the scaling factor is chosen to reduce the dependence on $N$. For example, $N_\ua=268$ for the experimental data plotted in Fig.~\ref{fig: density profile 2}, and is sometimes as high as $295$. 

Some of our calculations' inaccuracies may be mitigated by increasing the system size in the calculations with a uniform potential, including higher bands of the optical lattice, including finite-temperature corrections, or replacing the local density approximation with a BdG method which includes a spatially dependent potential.

\subsection{Scaled radii: Other parameters}

Here we explore the dependence of our results on the interaction strength and the number of atoms. In Figs.~\ref{fig: scaled radii}(a.i)-\ref{fig: scaled radii}(a.iv), we fixed $N_\ua=250$ and $a_s=-5360a_0$. In Figs.~\ref{fig: scaled radii}(b.i)-\ref{fig: scaled radii}(b.iv) and~\ref{fig: scaled radii}(c.i)-\ref{fig: scaled radii}(c.iv), we vary $N_\ua$ and $a_s$.

Figures~\ref{fig: scaled radii}(b.i)-\ref{fig: scaled radii}(b.iv) sets $N_\ua=100$, keeping $a_s=-5360a_0$ the same as in Figs.~\ref{fig: scaled radii}(a.i)-\ref{fig: scaled radii}(a.iv). The effect of reducing $N_\ua$ in our calculations is straightforward to understand -- the major effect is that it lowers the central chemical potential $\mu_c$. For $t<0.05\epsilon_B$, $\mu_c$ is reduced to a value below the chemical potentials where the FFLO phase is the ground state. Therefore, as can be observed from Figs.~\ref{fig: phase diags}(d)-\ref{fig: phase diags}(e), the gas always has a \SF\ core, surrounded by \NFP\ wings, and the FFLO phase is completely missing. For $t\geq0.05\epsilon_B$, the FFLO phase appears at large polarizations, but the size of the FFLO phase is smaller in Fig.~\ref{fig: scaled radii}(b.iv) than in Fig.~\ref{fig: scaled radii}(a.iv). Conversely, extrapolating this trend to increasing $N_\ua$ instead of decreasing $N_\ua$, $\mu_c$ will increase and the FFLO phase should be larger.

Based on the MF phase diagrams in Fig.~\ref{fig: phase diags}, we can predict that significantly increasing $N_\ua$ in MF theory, instead of decreasing $N_\ua$ as in Fig.~\ref{fig: scaled radii}(b), also leads to an important qualitative change in the arrangement of phases in the trap. For large $\mu_c$, a gas with $P_{\rm tube}\rightarrow0$ will have a 1D-like phase distribution, i.e., a FFLO core surrounded by \SF\ wings. While MF predicts that such 1D-like distribution of phases will appear only for large $N_\ua$, (e.g., $N_\ua > 700$ at $t=0.02\epsilon_B$ and $a_s=-5360a_0$), current experiments measure 1D-like profiles already for $N_\ua=250$ and $t\lesssim0.02\epsilon_B$. The system sizes required in our MF calculations to obtain a particle number high enough for a 1D-like phase distribution are prohibitively large.

Figures~\ref{fig: scaled radii}(c.i)-\ref{fig: scaled radii}(c.iv) set $a_s=-8610a_0$, keeping $N_\ua=250$ the same as in Figs.~\ref{fig: scaled radii}(a.i)-\ref{fig: scaled radii}(a.iv). The effect of changing $a_s$ in our calculations is more subtle. Changing the scattering length changes $\ell_B$, but we set $\ell_B=1$. However, changing $a_s$ and fixing $t/\epsilon_B$ requires appropriately changing the tunneling, which consequently changes $\omega_z$ due to the harmonic confinement provided by the 2D optical lattice. Thus, the only relevant difference between two plots in Figs.~\ref{fig: scaled radii}(a) and~\ref{fig: scaled radii}(c) with the same $t/\epsilon_B$ is the ratio $\hbar\omega_z/\epsilon_B$. This then leads to differences in $\mu_c/\epsilon_B$ and $h/\epsilon_B$ for a given polarization.

Figures~\ref{fig: scaled radii}(a.i)-\ref{fig: scaled radii}(a.iv) and~\ref{fig: scaled radii}(c.i)-\ref{fig: scaled radii}(c.iv) show a remarkable feature: the scaled radii for the same $t/\epsilon_B$ seem to be nearly identical, although the scattering lengths are different. This remarkable universal scaling of the scaled radii for different scattering lengths and equal $t/\epsilon_B$ was also observed in experiments~\cite{revelle20161d}. There seems to be no {\em a priori} reason for this universality. Nevertheless, we observe an apparent universal scaling in our calculations. One possible explanation could be the weak dependence of the scaled radii on $\hbar\omega_z/\epsilon_B$, as noted in Eqs.~\eqref{eqn: Rd} and~\eqref{eqn: perturbative Rd}.

\begin{figure}[t]\centering
\includegraphics[width=0.7\columnwidth]{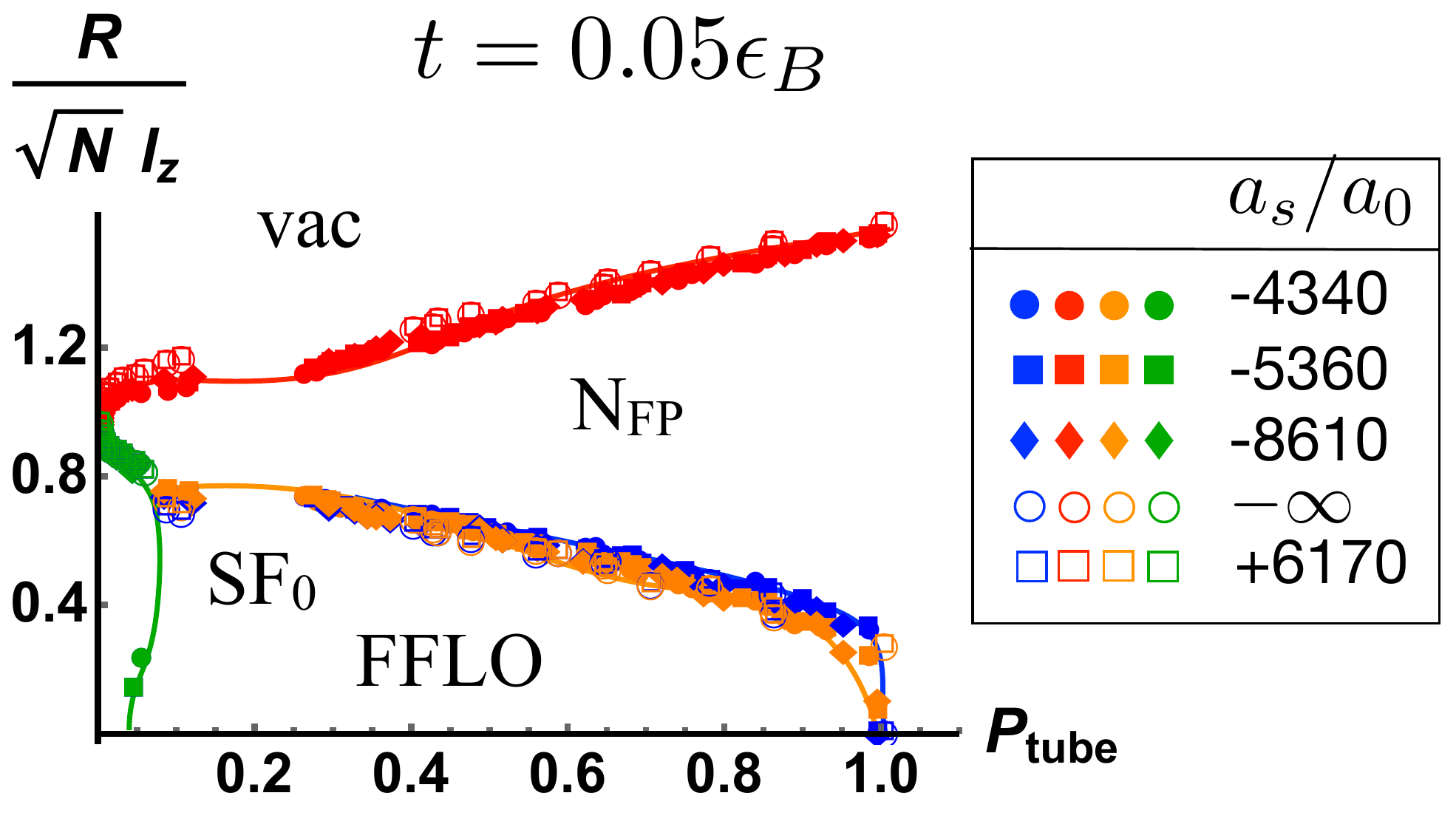}
\caption{(Color online) Apparent universality of scaled radii for different scattering lengths. The scaled radii at $a_s=-4340a_0, -5360a_0, -8160a_0, -\infty$, and $6170a_0$ are plotted as filled circles, filled squares, filled diamonds, open circles, and open squares, respectively. The lines guide the eye. The color coding for the different scaled radii $R_\da, R_d, R_{\rm SFi}$ and $R_\Delta$ is the same as that used in Fig.~\ref{fig: scaled radii}. The results for different scattering lengths do not collapse onto each other exactly, but show only a weak dependence on the scattering length.}
\label{fig: universality}
\end{figure}

We further analyze the universality of the scaled radii in Fig.~\ref{fig: universality}, where we plot the scaled radii for different scattering lengths at $t=0.05\epsilon_B$. The scattering lengths considered in Fig.~\ref{fig: universality} are the same as the scattering lengths that experiments set for the $^6$Li atoms by tuning the magnetic field~\cite{revelle20161d}. We observe that, although the scaled radii at different $a_s$ are nearly the same, they are not identical. In fact, a perturbative calculation of the scaled radii [Eq.~\eqref{eqn: perturbative Rd}] at small tunneling predicts a weak dependence on $\hbar\omega_z/\epsilon_B$ and thus on the scattering length, and our results are consistent with this.

\subsection{Onset of superfluidity}\label{subsec: p3d}
The 1D-ness or the 3D-ness of the phase distribution of a trapped gas is captured by plotting the critical polarization $P_{3D}$ at which the spin-balanced superfluid core shrinks to zero. If $P_{3D}=0$, then the gas does not have a \SF\ core as $P_{\rm tube}\rightarrow0$, and the gas is therefore 1D-like. If $P_{3D}\neq0$, then the gas is said to be 3D-like~\cite{revelle20161d}. The Bethe {\em Ansatz} shows that $P_{3D}=0$ at $t=0$~\cite{orso2007attractive, liao2010spin}. Consistent with this, experiments found that $P_{3D}=0$ for $t<0.02\epsilon_B$, and $P_{3D}>0$ for $t>0.02\epsilon_B$~\cite{revelle20161d}. They empirically associated this change in $P_{3D}$ with a crossover from 1D-like to 3D-like behavior of the gas at $t\sim0.02\epsilon_B$.

\begin{figure}[t!]\centering
\includegraphics[width=0.7\columnwidth]{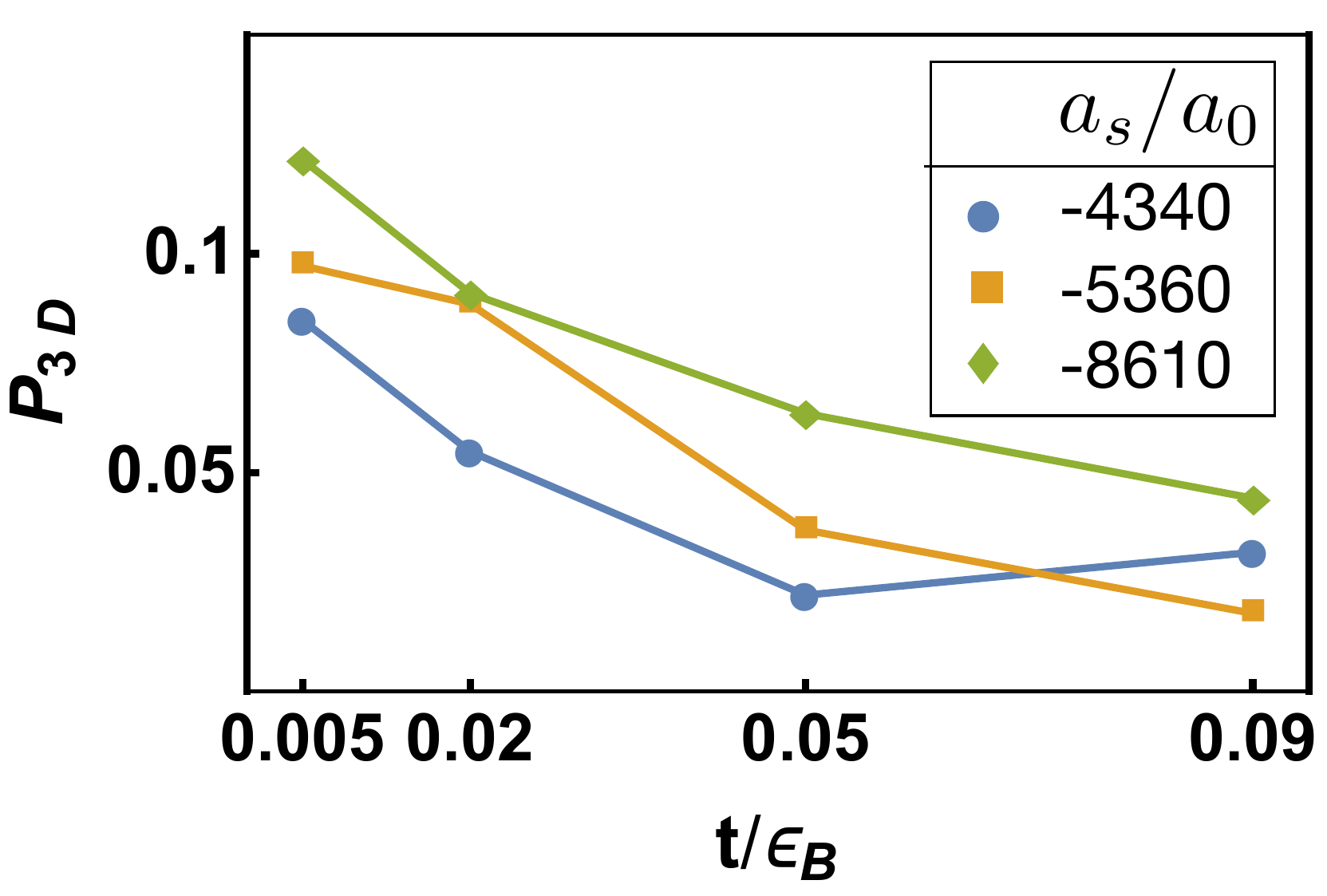}
\caption{(Color online) Critical polarization in MF theory for the onset of spin-balanced superfluidity at the center of the tubes, for different scattering lengths. The critical polarization for $a_s=-4340a_0, -5360a_0, -8160a_0$ is plotted as blue circles, orange squares, and green diamonds, respectively. The critical polarization in MF decreases with $t/\epsilon_B$, in contrast with experiments where it increases with $t/\epsilon_B$}
\label{fig: p3d}
\end{figure}

Figure~\ref{fig: p3d} plots $P_{3D}$ extracted from the MF scaled radii in Figs.~\ref{fig: scaled radii}(a) and~\ref{fig: scaled radii}(c). In contrast to experiments, we find that $P_{3D}>0$ for all $t/\epsilon_B$, and decreases with $t/\epsilon_B$. Notably, $P_{3D}\neq0$ even as $t\rightarrow0$, due to significant differences between our MF results and experimental measurements as well as the Bethe {\em Ansatz}.

The critical polarization is expected to change with particle number. With a large particle number in the trap, MF theory might lead to a 1D-like distribution of phases, giving $P_{3D}=0$ at small tunneling, and capture a crossover from $P_{3D}=0$ at small tunneling to $P_{3D}>0$ at large tunneling. But the particle numbers required for this are large. For example, MF predicts 1D-like behavior for $N_\ua > 700$ at $t=0.02\epsilon_B$ and $a_s=-5360a_0$, which is higher than the particle numbers $N_\ua\sim250$ in the present experiment.

\section{Experimental signatures}\label{sec: signature}
While experiments have been able to show the existence of the \NFP\ and \SF\ phases with {\em in situ} spin-sensitive density images, they have thus far not been able to prove the existence of the FFLO phase. Our numerical calculations indicate that the experimental measurements are consistent with having the FFLO phase in some regions of the cloud (see Fig.~\ref{fig: scaled radii}). Below, we argue that the experiments should be able to observe the FFLO phase with proper imaging techniques that can be implemented with current technology.

There are possibly two ways to experimentally observe the FFLO phase. The first method involves imaging the cloud after a time-of-flight expansion. Previously, theorists have predicted~\cite{kajala2011expansion, lu2012expansion} that the FFLO state shows clear peaks in the density after a time-of-flight expansion. The second method involves {\em in situ} imaging of small density oscillations. Below, we shed some light on the experimental requirements to measure these oscillations.

There are at least three questions to consider for imaging the oscillations {\em in situ} -- the magnitude and periodicity of oscillations in one tube, and the alignment of oscillations between different tubes. The LDA does not answer any of these questions, since we average over the oscillations at each chemical potential, but our calculations in a uniform potential can give some insight into the answers for one tube. The distance between density oscillations increases as $P_{\rm tube}$ decreases, and as we will see below, is within experimental imaging resolution only for small $P_{\rm tube}$. Therefore, we focus on the case of small $P_{\rm tube}$ here. For small $P_{\rm tube}$, the commensurate LO phase is more favorable than the FF or incommensurate LO phases.

In the commensurate LO phase, the excess $\ua$ spins are concentrated at domain walls. This causes the majority-spin density $n_\ua$ to peak at domain walls, and the minority-spin density $n_\da$ to dip at domain walls. The magnitude of the peak, relative to the background density away from the domain wall, is $O(\xi b^2)^{-1}$, where $\xi$ is the healing length and $b$ is the lattice spacing between the tubes. This magnitude is a large fraction of the background spin density $\overline{n_{\sigma,3D}} = \overline{n_\sigma}/b^2$, as, for example, evidenced in Fig.~\ref{fig: density profile 1}(a), and should be measurable in experiments. The number of excess spins per unit length in the commensurate LO phase is $n_\ua-n_\da$. Therefore, the average distance between the excess spins, i.e., the average distance between the density oscillations, is $1/(n_\ua-n_\da)$. For a typical experimental value of $\ell_B\sim O(200)$nm, and assuming that experiments can resolve distances larger than $4\mu$m, the density oscillations are resolvable if $(n_\ua-n_\da)\ell_B<0.05$. For the present experiments' parameters, MF calculations predict that this density difference can occur near the center of the trap for $P_{\rm tube}\lesssim0.05$ for $t/\epsilon_B\leq0.09$. More accurate values for the magnitude and periodicity of the oscillations can be obtained by doing a BdG calculation with a spatially dependent potential in the axial direction and a uniform potential in the transverse directions, instead of a uniform potential in all directions as we do in this paper.

Since our calculations assume that the chemical potential is uniform in the transverse directions, the densities in different tubes are identical, and therefore the density oscillations are always aligned. Generalizing the calculation to include a spatially dependent potential in the axial and transverse directions will shed light on the alignment of oscillations in different tubes. Our preliminary BdG calculations for two tubes with different potentials show that the oscillations in the two tubes are phase locked for sufficiently large tunneling, $t\gtrsim 0.05 \epsilon_B$. Doing a full 3D BdG calculation with spatially varying potentials in all directions is computationally expensive and subject to numerical difficulties such as getting stuck in local minima.

\section{Summary}\label{sec: summary}

We used Hartree-Fock Bogoliubov-de Gennes MF theory to calculate the phase diagram of a spin-imbalanced Fermi gas trapped in a 2D array of tunnel-coupled 1D tubes, and used the LDA to calculate the density profiles and scaled coordinates of the phase boundaries of this gas in an axially varying harmonic trap. We compared these results to experimental measurements of the density and phase boundaries~\cite{revelle20161d}, over a broad range of parameters.

Our calculations broadly agree with many aspects of these experimental measurements. We find density profiles and coordinates of the phase boundaries in a harmonic trap that are consistent with experimental measurements. We also reproduce the experimentally observed universal scaling of the scaled coordinates of phase boundaries onto one another for different scattering lengths, when $t/\epsilon_B$ is fixed.

However, our calculations show some discrepancies with the experimental measurements. While experiments measured a 1D-like distribution of phases in the trap, with a partially-spin-polarized core at the center of the trap at small polarizations and small tunneling, our calculations never produce such 1D-like behavior. Our calculations also yield an incorrect trend for the critical polarization for the onset of spin-balanced superfluidity. These inconsistencies between MF theory and experiments suggest beyond-mean-field effects play a significant role in the experiments. To capture these effects, it could be interesting to develop an approach starting from the exact $t=0$ Bethe {\em Ansatz} and incorporating weak tunneling between the tubes, as, for example, suggested by Ref.~\cite{zhao2008theory}. The 1D-ness of many of the experimental results suggests that such an approach, if it can be carried out, would be fruitful.

\section*{Acknowledgments}
This material is based upon work supported with funds from the Welch Foundation, Grants No. C-1872 and No. C-1133, the National Science Foundation Grants No. PHY-1848304 and No. PHY-1707992, and an Army Research Office MURI Grant No. W911NF-14-1-0003. K.R.A.H. thanks the Aspen Center for Physics, supported by the National Science Foundation Grant No. PHY-1066293, for its hospitality while part of this work was performed. B.S. thanks Erich Mueller and Shovan Dutta for useful conversations. We also thank Ben Olsen for his contributions to the experiment.

\bibliographystyle{apsrev4-1}
\bibliography{refs}
\end{document}